\documentclass[aps,prd,twocolumn,notitlepage,
superscriptaddress,
nofootinbib,floatfix]{revtex4-2}

\usepackage{comment}
\usepackage[utf8]{inputenc} 
\usepackage{amsfonts,amssymb, mathrsfs, amsmath, esint,bm,enumitem}
\usepackage{subcaption}
\usepackage{graphicx}
\usepackage{bm}
\usepackage{lineno}
\usepackage{amsmath}
\usepackage{float}
\allowdisplaybreaks

\usepackage{subcaption}
\usepackage{ragged2e}
\DeclareCaptionJustification{justified}{\justifying}
\makeatletter    
\renewcommand\onecolumngrid{
\do@columngrid{one}{\@ne}%
\def\set@footnotewidth{\onecolumngrid}
\def\footnoterule{\kern-6pt\hrule width 1.5in\kern6pt}%
}
\renewcommand\twocolumngrid{
        \def\footnoterule{
        \dimen@\skip\footins\divide\dimen@\thr@@
        \kern-\dimen@\hrule width.5in\kern\dimen@}
        \do@columngrid{mlt}{\tw@}
}%
\makeatother 

\usepackage{diagbox}
\usepackage{latexsym}
\usepackage{graphicx}
\usepackage[dvipsnames]{xcolor}
\usepackage{booktabs,multirow}
\usepackage{titlesec} 
\usepackage{physics}
\usepackage[normalem]{ulem}
\graphicspath{{./Figures/}}

\usepackage{hyperref}
\hypersetup{colorlinks, citecolor=ForestGreen, linkcolor=BrickRed, urlcolor=RedViolet}

\newcommand{\be}{\begin{equation}}
\newcommand{\ee}{\end{equation}}

\begin{document}

\title{Search for Peak Structures in the Stochastic Gravitational-Wave Background in LIGO-Virgo-KAGRA O1-O4a Datasets}

\author{C.-A.~Miritescu}
\affiliation{Institut de Física d'Altes Energies (IFAE) and\\ Barcelona Institute of Science and Technology (BIST), \\Campus UAB, Facultat Ciencies Nord, Bellatera, Barcelona, Spain}
\affiliation{Department of Physics, Universitat Autònoma de Barcelona (UAB), 08193 Bellaterra (Cerdanyola del Vallès), Barcelona, Spain}

\author{M.~Martinez}
\affiliation{Institut de Física d'Altes Energies (IFAE) and\\ Barcelona Institute of Science and Technology (BIST), \\Campus UAB, Facultat Ciencies Nord, Bellatera, Barcelona, Spain}  
\affiliation{Catalan Institution for Research and Advanced Studies (ICREA), Passeig de Lluís Companys, 23, Barcelona, Spain}

\author{O.~Pujolas}
\affiliation{Institut de Física d'Altes Energies (IFAE) and\\ Barcelona Institute of Science and Technology (BIST), \\Campus UAB, Facultat Ciencies Nord, Bellatera, Barcelona, Spain}


\begin{abstract}
We present a dedicated search for gravitational-wave backgrounds with nontrivial peak structures using data from the first three and the initial part of the fourth observing runs of the LIGO–Virgo–KAGRA network. The analysis is motivated by a variety of early-Universe models characterized by signals with multiple peaks. We introduce a model independent parameterization of double-peaked spectra based on the superposition of two normalized broken power laws and perform a Bayesian inference study using the LIGO-Virgo-KAGRA isotropic cross-correlation data. While no statistically significant evidence for a multi-peak background is found, the analysis provides constraints on the inter-peak slopes in correlation with the signal amplitude. These results exhibit LIGO-Virgo-KAGRA's ability to probe signals beyond a single peak structure and establish a foundation for future targeted searches for nontrivial gravitational waves background spectral shapes in future observing runs and the advanced detector era.   
\end{abstract}

\maketitle

\section{Introduction}

Stochastic gravitational wave backgrounds (SGWBs), arising from the superposition of unresolved gravitational wave (GW) signals \cite{Christensen_2018},  provide a powerful probe of both the astrophysical sources populations \cite{de_Araujo_2000, Regimbau_2006, Regimbau_2011, Abbott_2016, Jenkins_2019, ebersold2025} and the fundamental processes in early-Universe cosmology \cite{O4cosmological, Caprini_2018}. 

While recent results from several pulsar timing array (PTA) collaborations have provided compelling evidence for the presence of a nanohertz-frequency stochastic gravitational wave background \cite{NANOGrav2023, EPTAInPTA2023, PPTA2023, CPTA2023}, a SGWB can, in principle, be also probed using current ground-based gravitational wave detectors, such as the LIGO-Virgo-KAGRA (LVK) network \cite{Allen1999, Maggiore2000, galaxies10010034}. The LVK detector network operates in the $10–1000$ Hz frequency band. The LVK collaboration has performed extensive searches for isotropic SGWBs assuming power-law spectra \cite{O2, O3, O4isotropic}, with the most recent upper limit for such a signal being $\Omega_{ref} = 2 \times 10^{-9}$ at a $95\% $ confidence level. Several searches for an anisotropic gravitational wave background have also been conducted \cite{KAGRA:2021mth, KAGRA:2021rmt, LIGOScientific:2025bkz}. The search  SGWBs is further motivated by recalling that PTAs are sensitive to primordial sources at the QCD epoch and correspondingly LVK is to earlier epochs when the temperature was as high as around $10^8\,$GeV \cite{Maggiore2000,Caprini:2024ofd}.

It is important to note that the sources producing backgrounds in different frequency ranges can be different. For example, the astrophysical PTA background is primarily thought to be produced by inspiral supermassive black hole binaries (SMBHB) \cite{Sesana_2013, Sesana_2008, Rajagopal}, while the astrophysical component of the LVK background is thought to be sourced mostly by inspiral stellar mass black holes \cite{Rosado_2011,Zhu_2011, PhysRevD.85.104024, Abbott_2016, Zhu_2013}. Additional astrophysical contributions for the background may arise from core-collapse supernovae \cite{Buonanno, Sandick, core}, stellar core-collapse \cite{crocker, crocker2}, rotating neutron stars \cite{Lasky, Ferrari, Rosado, wu}, or boson clouds around black holes \cite{Brito, Brito_2017, Tsukada, Boson}. 

Beyond astrophysical processes, a stochastic background can also originate from cosmological sources, encoding signatures of the very early Universe. These signals, including those produced during the dark ages prior to the Cosmic Microwave Background (CMB) decoupling, constitute a unique observational window into primordial cosmology, probing epochs and energy scales unreachable by any other means \cite{maggiore1998}. These include phenomena such as gravitational waves generated during inflation \cite{Barnaby, Barnaby_2011, Maleknejad_2016, Thorne_2018, Bartolo_2016, badger, ABBOTT1984541}, first-order phase transitions \cite{Caprini2016, Mazumdar_2019, causality, Romero_2021, Caprini:2024ofd, Caprini:2024gyk}, cosmic strings \cite{Kibble:1976sj, Damour, Siemens, auclair2020probing}, primordial black holes \cite{Mandic2016, Wang_2018, Andres-Carcasona:2024wqk, Raidal_2017,BAGUI2022101115, Braglia_2021, Mukherjee_2021, Ferrer:2018uiu, Gelmini:2023ngs, Gouttenoire:2023gbn}, domain walls annihilation \cite{Oliveira_2005, ferreira, Gelmini:2022nim, Ferreira:2022zzo, Ferreira:2024eru, Notari_2025} and other high-energy mechanisms operating at large energy scales \cite{O4cosmological, Caprini_2018}.

Depending on the underlying dynamics, the same cosmological mechanism can produce GW spectra spanning several order of magnitude in frequency, allowing multi-band detection prospects across different experiments. For instance, a background generated by cosmic strings may contribute at nanohertz frequencies probed by PTAs, millihertz frequencies accessible to LISA, and even at the hundreds-of-hertz range observable by the LVK network \cite{auclair2020probing, galaxies10010034}). The complementarity between frequency bands is therefore crucial: low-frequency detectors probe long-lived, large-scale processes such as SMBHB inspirals or cosmic string networks, while ground-based interferometers explore the high-frequency regime where compact object mergers and certain cosmological relics may leave observable imprints. A coordinated multi-band effort thus provides a powerful framework to disentangle astrophysical and cosmological contributions to the SGWB, potentially revealing new physics and offering insights into the Universe’s earliest moments.

Standard searches typically assume a single-peaked or power-law spectrum \cite{O1,O2, O3, O4isotropic, O4cosmological}; however, an increasing number of cosmological models predict multi-peaked structures in the SGWB energy density spectrum, driven by complex dynamics in the early Universe. In particular, double-peaked SGWB spectra can naturally emerge in scenarios involving multistep first-order phase transitions \cite{Morais:2019fnm,Bigazzi_2021, zhao, Aoki_2022, Benincasa:2022elt, Cao:2022ocg, Chen:2025ksr, cbgr-w9cb}, hybrid inflation \cite{Dufaux:2008dn, Guzzetti:2016mkm}, oscillons formed after inflation \cite{Liu_2018, Zhou:2013tsa, Hiramatsu:2020obh}, scalar induced GWs \cite{ Domenech:2021ztg, Cai_2019, Roshan:2024qnv},  GWs induced by both adiabatic and isocurvature modes \cite{Bhaumik_2022} or an early stage of matter domination \cite{Fernandez:2023ddy, Dalianis:2024kjr}. These models predict features in the GW spectrum that may span different frequency regimes — sometimes within the detection band of ground-based interferometers.

The LVK collaboration has performed extensive searches for isotropic SGWBs assuming power-law spectra \cite{O2, O3, O4isotropic}, but dedicated analyses targeting double-peak spectral morphologies are still in their early stages \cite{Yu:2022xdw, Miritescu:2023ruv}. Identifying such signals would offer compelling evidence of complex high-energy processes in the early Universe and help distinguish between competing beyond-Standard-Model scenarios.

Moreover, given the relatively narrow frequency coverage of ground-based detectors compared to space-based missions or PTAs, optimized detection strategies are crucial for identifying non-trivial spectral shapes within the LVK data.

In this paper, we develop and implement search techniques for identifying double-peak SGWB signals in LVK data. Using data from the first three observing runs and from the first part of the fourth observing run performed by LVK, we assess the sensitivity of the network to a range of benchmark double-peak models and place constraints on their amplitudes. This work represents a first step toward a systematic program to search for complex cosmological GW backgrounds with current and future ground-based detectors.

This paper is structured as follows. In Section II, we describe the methodology used to convert LVK strain data into a cross-correlation estimate of the GW energy-density spectrum. Section III introduces the double-peak parametrization used to model potential signals. Section IV details the analysis method, including the Bayesian parameter estimation. We conclude in Section V with a presentation of the results and a discussion of their implications.

\section{Methodology}

The gravitational wave background is characterized by its energy density per logarithmic frequency interval, divided by the critical energy density of the Universe:
\[\Omega_{GW}(f) = \frac{f}{\rho_c}\frac{d\rho_{GW}}{df},\]
where $d\rho_{GW}$ is the gravitational waves energy density in the frequency interval $f$ to $f + df$ and $\rho_c$ is the critical energy density of a flat Universe.

In the absence of correlated noise, the cross-correlated strain data between pairs of LIGO-Virgo-KAGRA gravitational-wave detectors can be used as en estimator for the gravitational waves background spectrum \cite{vuk, O1, O2, O3, O4isotropic}. This estimator is computed by following the steps below, according to \cite{Allen1999}.

The cross-spectral density (CSD) is derived from the Fourier-transformed strain data $\Tilde{s}(f)$, corresponding to an observation period $T$, recorded by two geographically separated detectors $I$ and $J$.
\[C_{IJ} = \frac{2}{T}\Tilde{s}^*_{I}(f)\Tilde{s}_{J}(f)\]

The GW energy density estimator can then be computed as:
\[\Omega_{GW}(f) = \frac{\text{Re}[C_{IJ}(f)]}{\gamma_{IJ}(f)S_0(f)}\]
where $\gamma_{IJ}(f)$ is the overlap reduction function (ORF), a geometric parameter which quantifies the frequency-dependent loss of coherence in the cross-correlation of signals from two spatially separated interferometers \cite{Finn_2009, Romano2017}, and $S_0(f)$ is used to convert units of GW strain power into fractional energy density.

Its variance can be calculated in the following way:
\[\sigma_{IJ}^2(f) = \frac{1}{2T\,\Delta f}\,\frac{P_I(f)\,P_J(f)}{\gamma_{IJ}^2(f)\,S_0^2(f)}\]
where $P_I(f)$ is the one-sided power spectral density for detector I, and $\Delta f$ is the frequency resolution employed in the analysis.

To constrain the parameters of the cosmological model, we adopt a Bayesian inference framework~\cite{Christensen_2022}, utilizing data from the first LIGO--Virgo observing runs (O1), the second LIGO--Virgo observing run (O2), the third LIGO--Virgo--KAGRA observing run (O3), and the first part of the fourth LIGO--Virgo--KAGRA observing run (O4a) \cite{O4isotropic}. For brevity, we will refer to this dataset as the O1–O4a LVK data.\footnote{While the data are part of the LVK collaboration, our analysis uses LIGO data for the O1–O4a runs combined with Virgo data solely for the O3 run.} We follow the analysis procedure outlined in \cite{vuk}. Assuming that the cross-correlation estimator $C_{IJ}(f)$ follows a Gaussian distribution, the likelihood function can be expressed as:

\begin{equation} 
\begin{split} 
&p(C_{IJ}(f) \mid \theta) \propto \\ 
&\exp\left[ -\frac{1}{2} \sum_f \frac{\left( C_{IJ}(f) - \Omega_{\mathrm{GW}}(f, \theta) \right)^2} { \sigma_{IJ}^2(f) } \right], 
\end{split} 
\label{eq:likelihood} 
\end{equation}
where the data are taken from detectors $I$ and $J$, and $\sigma_{IJ}^{2}(f)$ denotes the variance introduced above. Both $C_{IJ}(f)$ and $\sigma_{IJ}^2(f)$ are provided as data products by the LIGO--Virgo--KAGRA Collaboration’s isotropic gravitational-wave background (GWB) search~\cite{O4isotropic}, derived directly from the detector strain data using the \textit{pygwb} pipeline \cite{Renzini_2023, Renzini2024}. The isotropic analysis assumes the absence of correlated noise between detectors $I$ and $J$, such as that induced by correlated magnetic fields, as discussed in \cite{Meyers_2020, Kamiel2023, janssens2025coherentinjectionmagneticnoise}.

The function $\Omega_{GW}(f\,,\theta)$ denotes the target model of the search, characterized by the set of parameters $\theta$.

\section{Parameterization}
The total SGWB power spectrum is expressed here as the sum of the compact binary coalescence signals (CBCs) and the double-peaked cosmological contribution (Cosmo): 
 \[ \Omega_{GW}(f, \theta) = \Omega_{CBC} (f, \theta_{CBC}) + \Omega_{Cosmo} (f, \theta_{Cosmo})\]

The CBC contribution used is the usual power law expected from astrophysical sources which was used in previous analysis \cite{O1, O2, O3,O4isotropic}:
\begin{equation} \label{eq:CBC}
\Omega_{CBC}(f, \theta_{CBC}) = \Omega_{ref}\left(\frac{f}{f_{ref}}\right)^{\alpha}
\end{equation}

The set of parameters $ \theta_{CBC}$ is: $\Omega_{ref}$ - the amplitude of the astrophysical signal, $\alpha$ - the power of the dependency, $ f_{ref}$ - the reference frequency, usually chosen to correspond to the highest sensitivity of the detector.

The double-peaked signal $\Omega_{Cosmo}$ is modeled as a sum of two individual, normalized broken power laws given by:
\begin{equation}
\Omega_{BPL}\left(f\right) = \Omega_*S\left(f\right)
\end{equation}
 where $\Omega_{*}$ is the amplitude of the signal and 
\begin{equation}
\begin{split} 
&S\left(f\right)=\left(1 - \frac{n_1}{n_2}\right)^{\frac{n_1 - n_2}{\Delta}} \times\\
&\left(\frac{f}{f_{*}}\right)^{n_1}\left(1 - \frac{n_1 }{n_2}\left(\frac{f}{f_{*}}\right)^{\Delta}\right)^{\frac{n_2 - n_1}{\Delta}}  
\end{split}
\end{equation} 
For a broken power law function, $n_1$ is the power describing the behaviour of the function at low frequencies, while $n_2$ describes the behaviour at high frequencies. $\Delta$ is a smoothing factor: the larger the value of $\Delta$, the sharper the transition between the two power slopes. $f_*$ is the frequency where the transition between the two powers occurs. This shape has been previously used in some O1+O2+O3 LVK data studies in \cite{Romero_2021}, \cite{Miritescu:2023ruv}, and \cite{Badger:2022nwo}, with some variations in the prefactors of the function. In this work, we focus on this particular expression of $S(f)$ due to its property to be normalized at $f=f_{*}$: $S(f_*)=1$ and its peak occurring precisely at $f_{*}$. 

The detailed expression for $\Omega_{Cosmo}$ will then be:
\begin{equation} 
\begin{split} 
&\Omega_{Cosmo}(f, \theta_{Cosmo}) = \\
&\Omega_{*1} \left(1 - \frac{n_1}{n_2}\right)^{\frac{n_1 - n_2}{\Delta_1}}\left(\frac{f}{f_{*1}}\right)^{n_1} \times\\
&\left(1 - \frac{n_1}{n_2}\left(\frac{f}{f_{*1}}\right)^{\Delta_1}\right)^{\frac{n_2 - n_1}{\Delta_1}} \\
& + \Omega_{*2} \left(1 - \frac{n_3}{n_4}\right)^{\frac{n_3 - n_4}{\Delta_2}}\left(\frac{f}{f_{*2}}\right)^{n_3} \times\\
&\left(1 - \frac{n_3}{n_4}\left(\frac{f}{f_{*2}}\right)^{\Delta_2}\right)^{\frac{n_4 - n_3}{\Delta_2}}\label{eq:spectrumdp}
\end{split}
\end{equation}

The set of parameters $\theta_{Cosmo}$ will be represented by: $\Omega_{*1}$, $f_{*1}$, $\Omega_{*2}$, $f_{*2}$, $n_1$, $n_2$, $n_3$, $n_4$, $\Delta_{1}$, $\Delta_2$. 

\begin{figure}[h]
\centering
\includegraphics[width =1\linewidth]{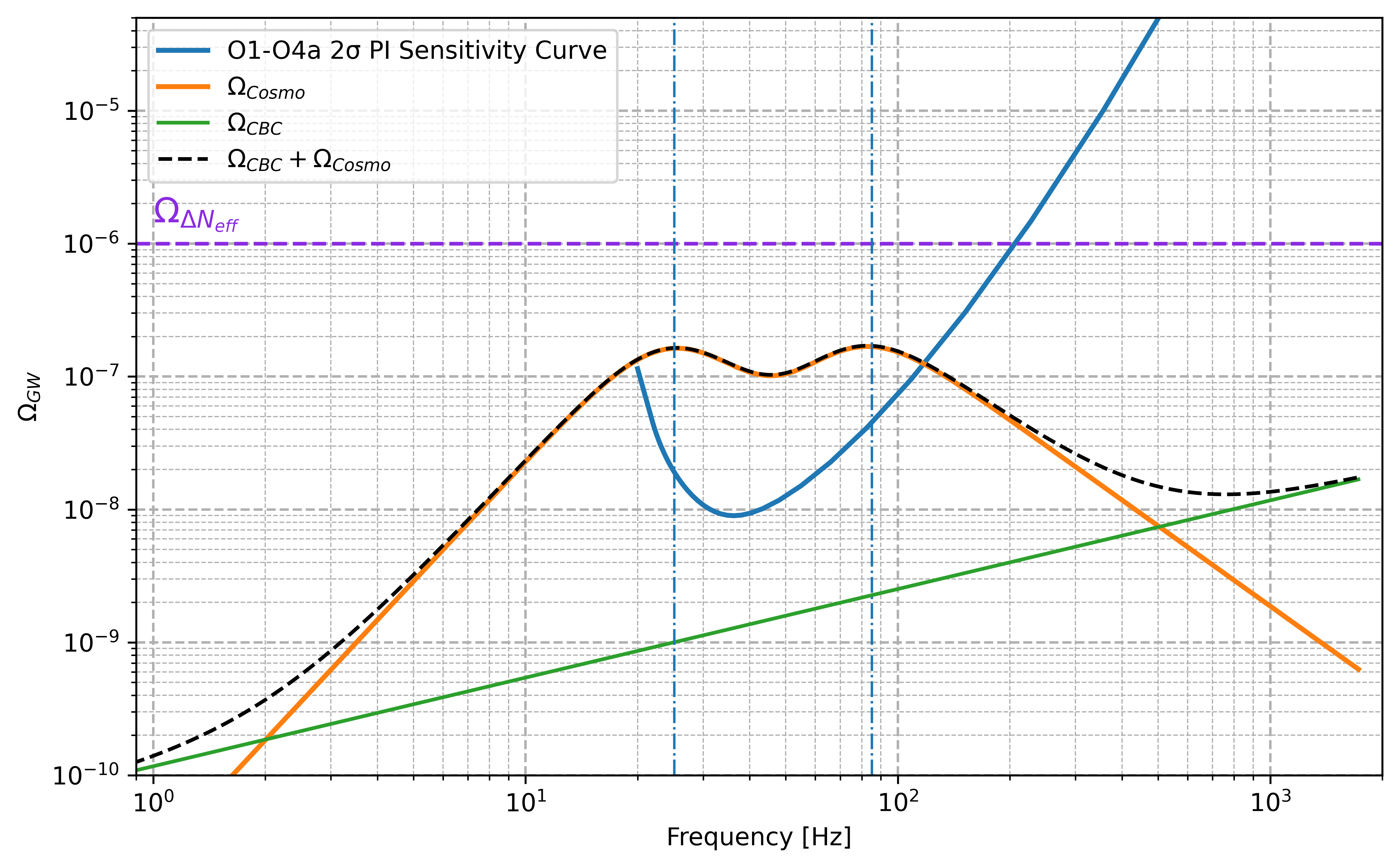}
\caption{Example of the double-peaked signal, where both peaks are located in the sensitivity range and above the power-law integrated sensitivity curve of the detector. The blue dashed lines indicate the peaks' frequencies. The horizontal purple dashed line marked by $\Omega_{\Delta N_{eff}}$ indicates the constraints coming from CMB and BAO analyses \cite{Caprini_2018}. }
\label{fig:both_i}
\end{figure}

\begin{figure}[h]
\centering
\includegraphics[width =1\linewidth]{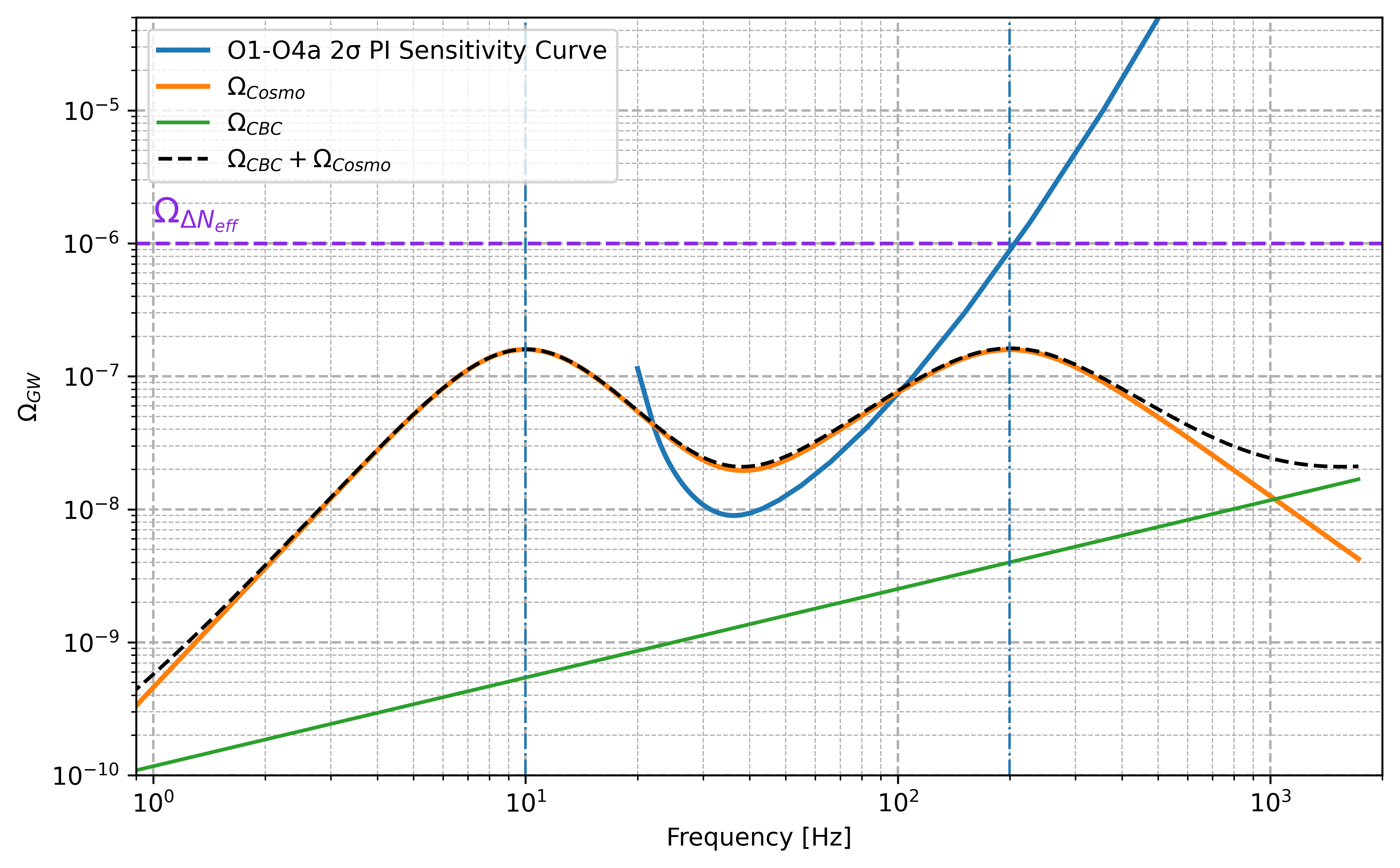}
\caption{Example of the double-peaked signal where both peaks are located outside the sensitivity range of the detector, but the valley in between the peaks is above the power-law integrated sensitivity curve. The blue dashed lines indicate the peaks' frequencies. The horizontal purple dashed line marked by $\Omega_{\Delta N_{eff}}$ indicates the constraints coming from CMB and BAO analyses \cite{Caprini_2018}.}
\label{fig:both_o}
\end{figure}

The combination of these functions gives rise to the $\Omega_{GW}$ signal shape presented in Figs. \ref{fig:both_i} and  \ref{fig:both_o}. 

In Fig. \ref{fig:both_i}, the parameters were chosen to demonstrate the feasibility of detecting such a signal, ensuring that both peaks appear within the sensitive band of the detector and above the 2-sigma power-law integrated sensitivity curve \cite{thrane, O4isotropic}. In Fig. \ref{fig:both_o}, the parameters were selected such that the valley between the peaks is in the sensitivity range and above the 2-sigma power-law integrated sensitivity curve \cite{thrane, O4isotropic}.

At this stage, the number of free parameters of the CBC + Cosmological model is 13, representing a substantial parameter space and a computational challenge.

\section{Analysis}

In order to extract information from the Bayesian analysis, we need to set some of the parameters whose best values have been established either through physics arguments or by means of numerical simulations. 

The CBC background is described by the usual values for $\alpha=\frac{2}{3}$ \cite{Regimbau_2011, Callister2016} and $f_{ref} = 25$ Hz, corresponding to the most sensitive frequency for SGWB searches and as used in previous LVK analyses \cite{O1, O2, O3, O4isotropic}. This approximation is valid only for the current LVK sensitivity range. For future detectors, the model will need to be updated. 

The ascending slope of the first peak is set to $n_1=3$ as motivated by causality considerations \cite{causality}. Thus we are left with 10 free parameters, which we use to perform a wide search and for which we need to select appropriate priors. These priors are presented in Table \ref{table:tab1}.

\begin{table}[h!]
\caption{\label{table:tab1}Priors selected for the wide search.}
\begin{ruledtabular}
\begin{tabular}{ccc}
Param. & Prior type &Prior range\\
 \hline
 \\
 $\Omega_{ref}$   & LogUniform    &$(10^{-11},10^{-6})$\\
$\Omega_{*1}$&   LogUniform  & $(10^{-11},10^{-6})$ \\
$f_{*1}$&   LogUniform  & $(10^{-3}$ Hz$,10^{3}$ Hz$)$\\
 $\Omega_{*2}$    &LogUniform & $(10^{-11},10^{-6})$\\
 $f_{*2}$&   LogUniform  & $(10^{-3}$ Hz$,10^{3}$ Hz$)$\\
 $n_2$& Uniform  &$(-6, 0)$ \\
 $n_3$& Uniform  &$(0, 6)$ \\
 $n_4$&Uniform  & $(-6, 0)$\\
 $\Delta_1$ & Uniform& $(1,8)$\\
 $\Delta_2$ & Uniform& $(1,8)$
\end{tabular}
\end{ruledtabular}
\end{table}

The amplitude of the CBC stochastic background emission and the two peak amplitudes and the frequencies were sampled in a $log-uniform$ fashion, while the exponents were sampled in a $linear-uniform$ way. The amplitudes range was chosen to contain the most recent upper limit estimations \cite{O4isotropic}. The frequency range was chosen to correspond to the peak sensitivity of the LIGO-Virgo detectors. To avoid coincident peaks and eliminate the possibility of interchanging the two frequencies, an additional constraint was applied to enforce an ordering between $f_{*1}$ and $f_{*2}$: $f_{*2}>f_{*1}$. This effect is visible in the $log(f_{*1})$ vs. $log(f_{*2})$ parameter space in Fig. \ref{fig:corner_wide} in appendix A. We then perform a Bayesian analysis using the available LVK data and the Bilby library \cite{bilby}.

The posterior distributions of the parameters for the Bayesian search using the priors described in the table above are presented in Fig. \ref{fig:corner_wide} in appendix A.

For the amplitudes, we obtain the following upper limits, with a $95\%$ confidence level:  $\Omega_{ref} = 2.9 \times 10^{-9}$, $\Omega_{*1} = 4.6 \times 10^{-7}$ and $\Omega_{*2} = 2 \times 10^{-7}$. These values are compatible with the results obtained in the LVK O4 stochastic gravitational waves background search paper \cite{O4isotropic}, where the numerical value for $\Omega_{ref}$ was found to be $\Omega_{ref}= 2 \times 10^{-9} $. For most of the exponents, we obtain flat posteriors, indicating no preference for any particular value. The exceptions are $n_2$ and $n_3$, whose behavior will be investigated further. 

The Bayes factor calculated between a model containing only noise and the model composed of the \textit{CBC + Cosmological Double Peak} gravitational waves stochastic background has the value: 
\[logB^{CBC+Cosmo}_{Noise} = -1.32\]

We can conclude for this search that there is no evidence to support this model over a purely Gaussian noise one.

The power indices $n_4$, $\Delta_{1}$ and $\Delta_{2}$ have completely flat posteriors, and thus for a benchmark search we set them to have the values: 
\[n_4 = -2 \]
\[\Delta_{1} = \Delta_{2}=4\]

The choice for $n_4$ is motivated by the fact that a finite-size, finite-duration phase-transition source, for example, cannot generate arbitrarily short-wavelength gravitational waves, and thus the spectrum is required to fall at high frequencies. The values for $\Delta$s were chosen to ensure that the transition between the two power-law regimes is neither unrealistically sharp nor excessively smooth.
We reinforce the ordering between the two peak frequencies by explicitly selecting non-overlapping priors (Table \ref{table:tab2}). Here, the analysis targets the data sensitivity to fluctuations within the valley between the peaks.
\begin{figure*}
    \centering
    \begin{subfigure}[t]{0.32\textwidth}
        \centering
        \includegraphics[width=\linewidth]{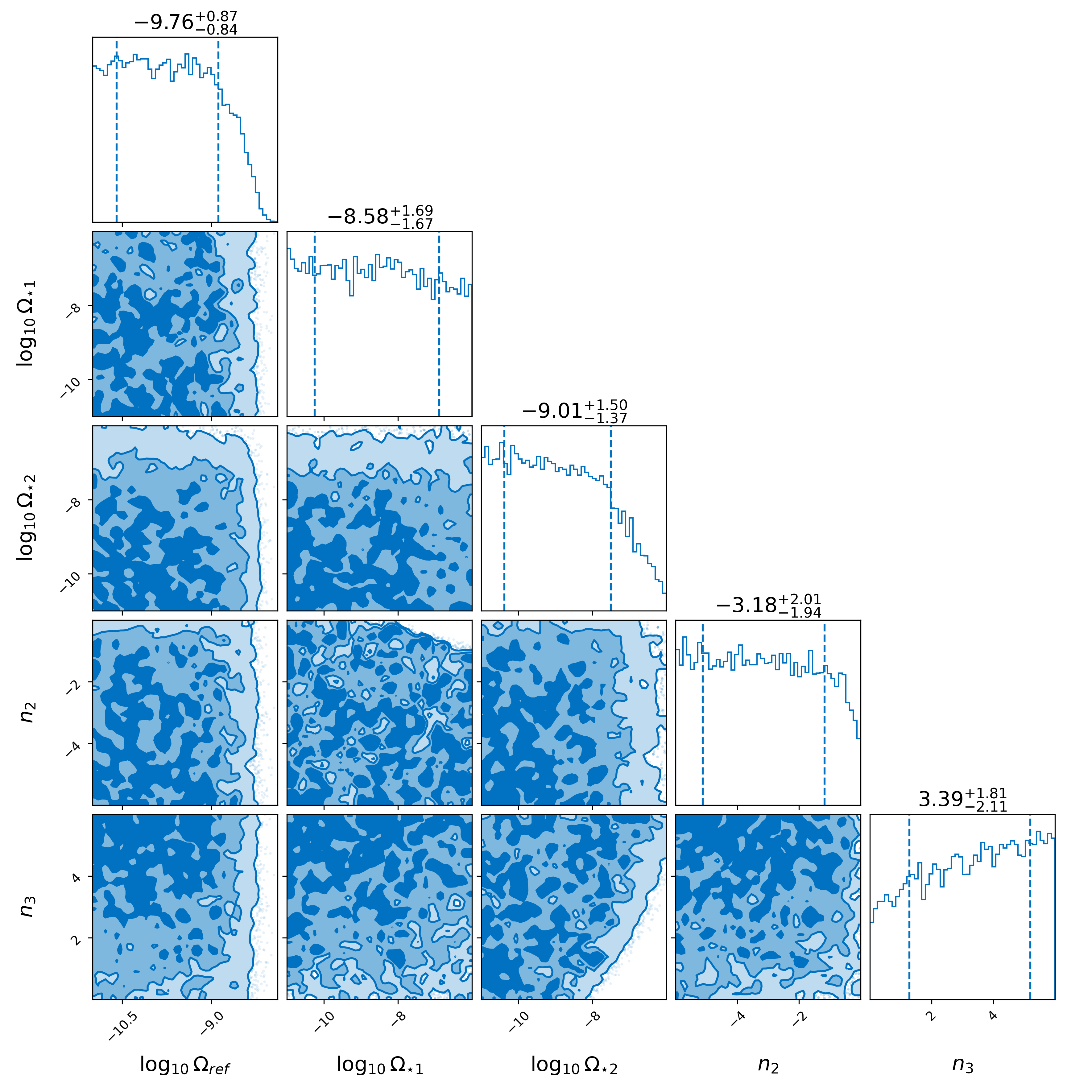} 
        \caption{$f_{*1}=0.1$ Hz} \label{fig:timing1}
    \end{subfigure}
    \hfill
    \begin{subfigure}[t]{0.32\textwidth}
        \centering
        \includegraphics[width=\linewidth]{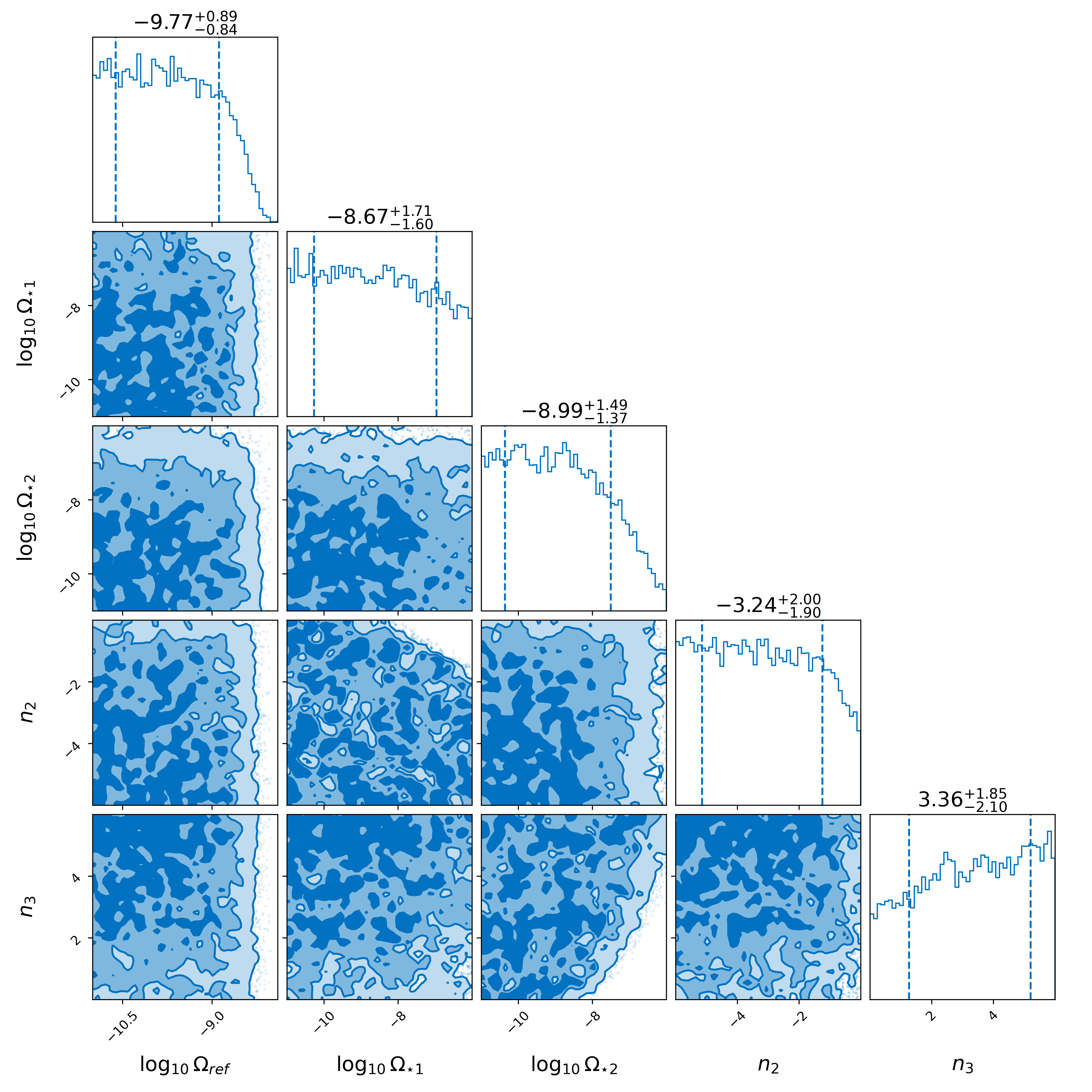} 
        \caption{$f_{*1}=1$ Hz} \label{fig:timing2}
    \end{subfigure}
    \hfill
    \begin{subfigure}[t]{0.32\textwidth}
    \centering
        \includegraphics[width=\linewidth]{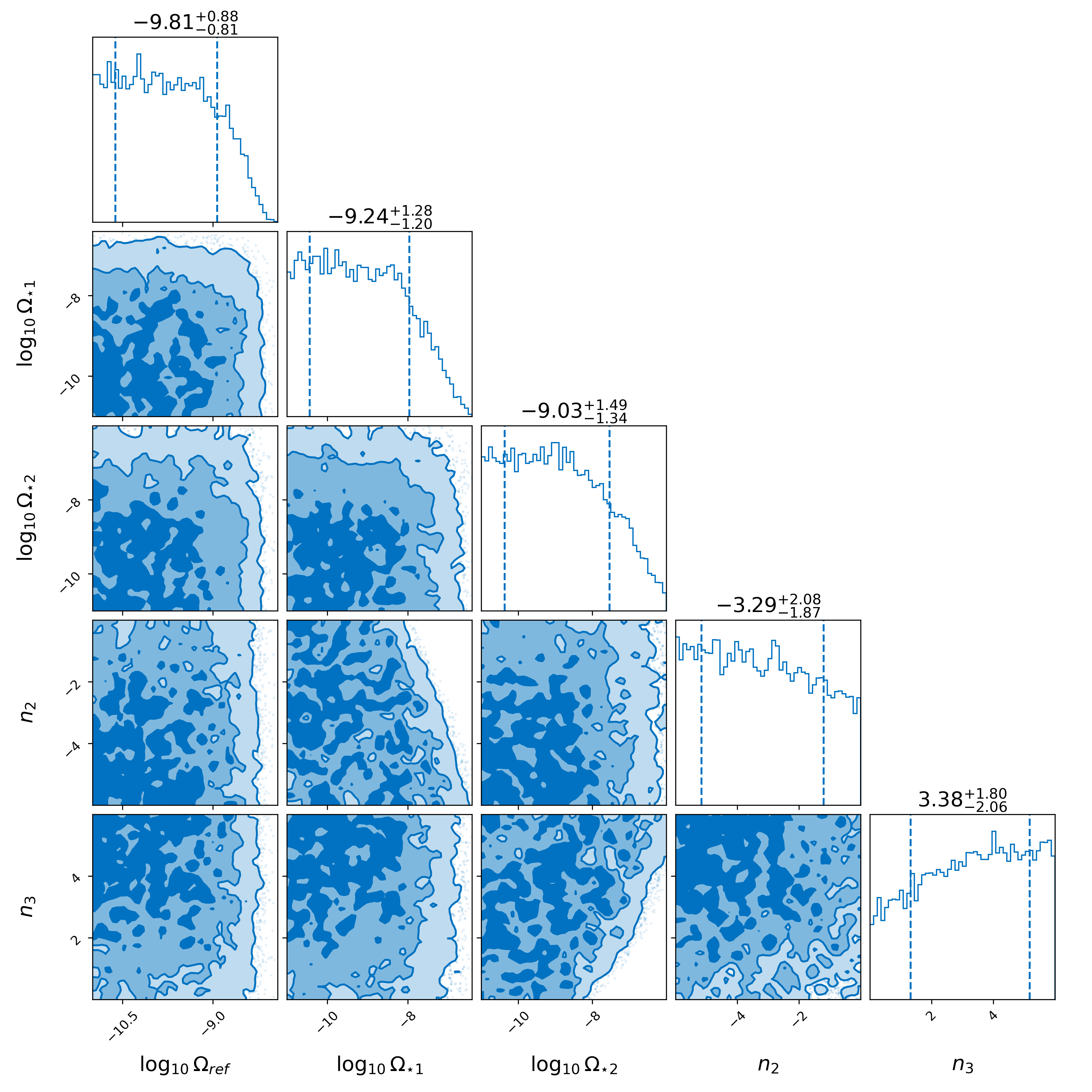} 
        \caption{$f_{*1}=10$ Hz} \label{fig:timing3}
    \end{subfigure}
    \caption{Posterior distributions obtained when fixing the location of the second peak $f_{*2} = 300$ Hz. As the frequency $f_{*1}$  varies, a region in the $n_2$ vs. $\Omega_{*1}$ parameter space can be excluded.} \label{fig:n2}
\end{figure*}

\begin{figure*}
    \centering
    \begin{subfigure}[t]{0.32\textwidth}
        \centering
        \includegraphics[width=\linewidth]{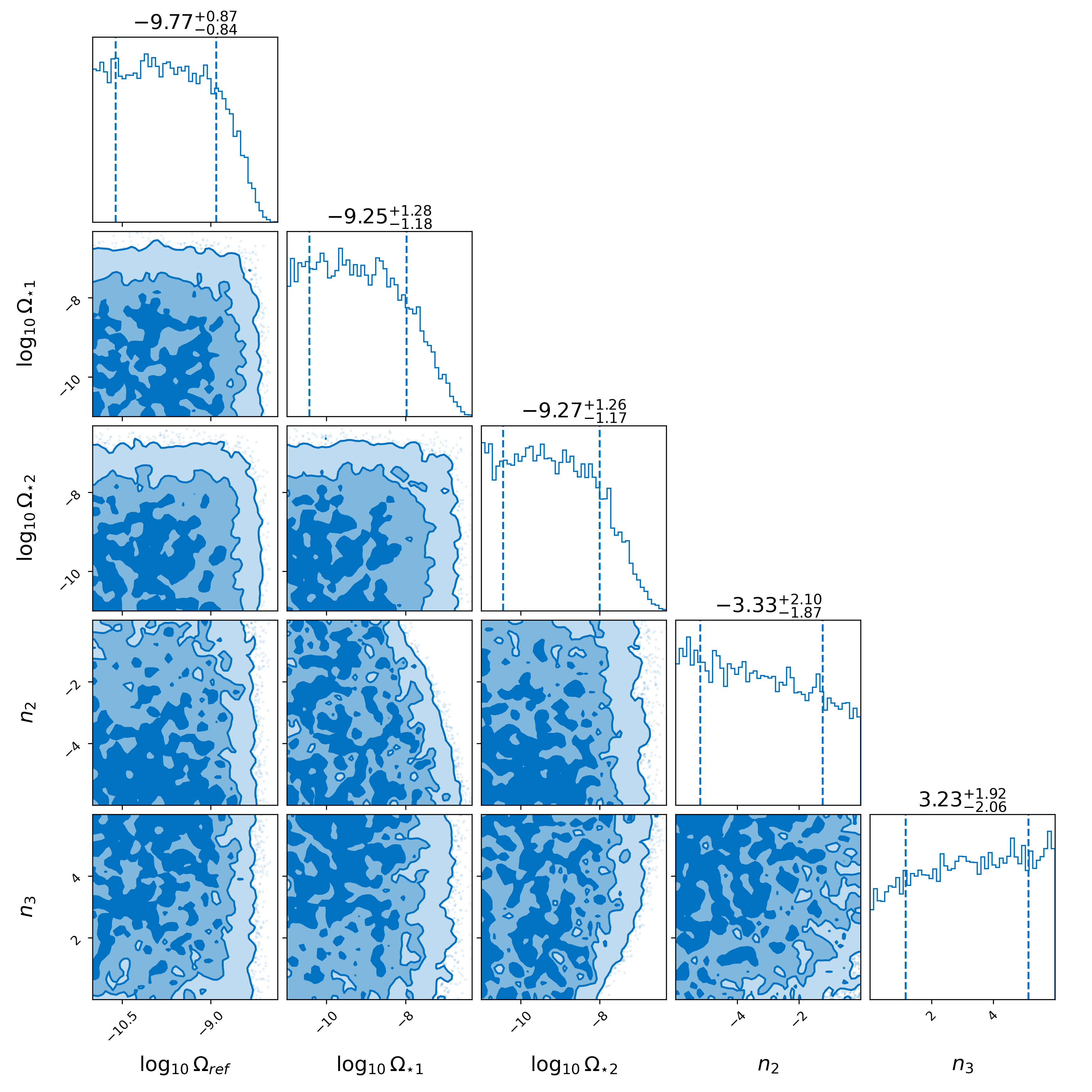} 
        \caption{$f_{*2}=200$ Hz} \label{fig:timing1}
    \end{subfigure}
    \hfill
    \begin{subfigure}[t]{0.32\textwidth}
        \centering
        \includegraphics[width=\linewidth]{10-300.png} 
        \caption{$f_{*2}=300$ Hz} \label{fig:timing2}
    \end{subfigure}
    \hfill
    \begin{subfigure}[t]{0.32\textwidth}
    \centering
        \includegraphics[width=\linewidth]{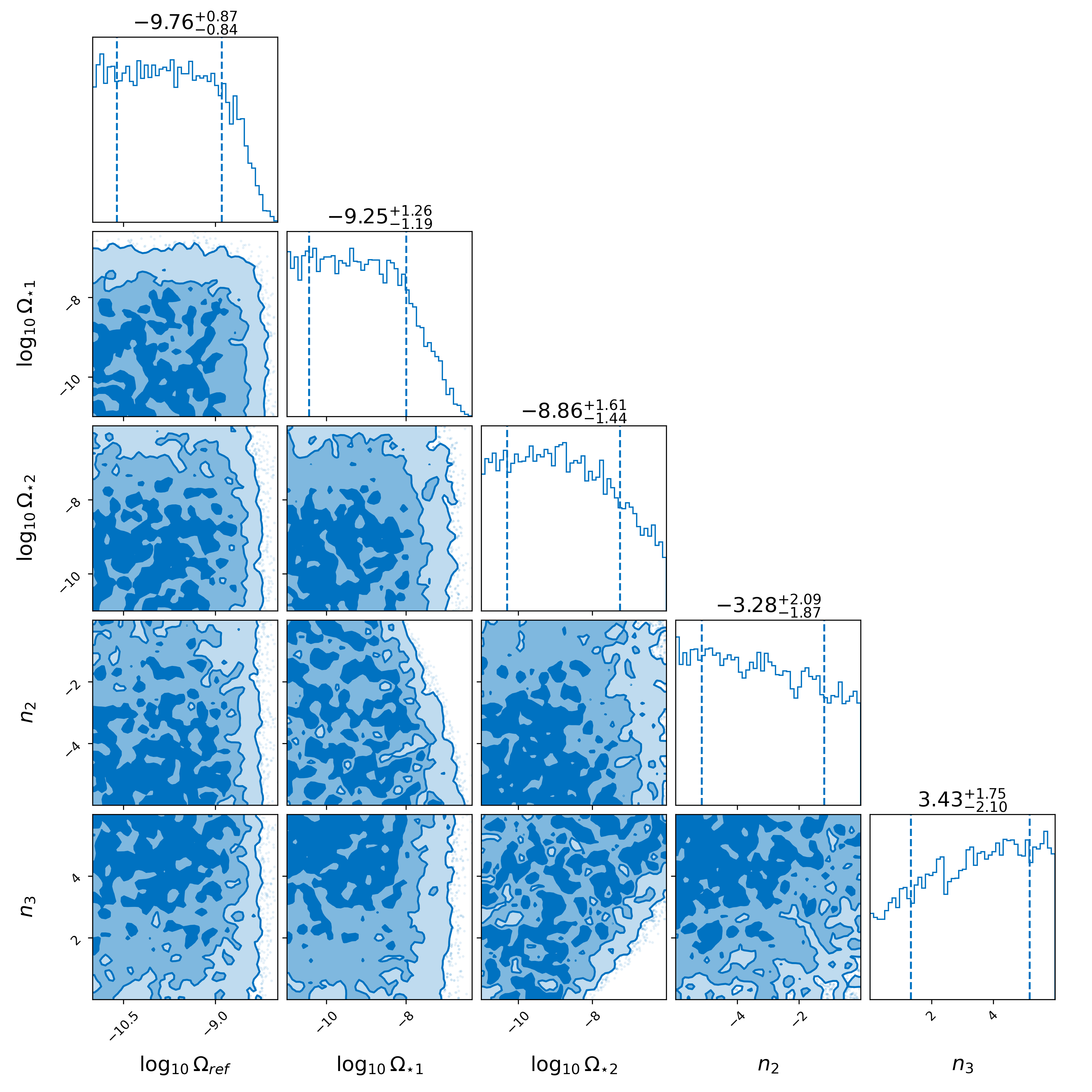} 
        \caption{$f_{*2}=400$ Hz} \label{fig:timing3}
    \end{subfigure}
    \caption{Posterior distributions obtained when fixing the location of the first peak $f_{*1} = 100$ Hz. As the frequency $f_{*2}$  varies, a region in the  $n_3$ vs. $\Omega_{*2}$ parameter space can be excluded.} \label{fig:n3}
\end{figure*}

\begin{table}[h!]
\caption{\label{table:tab2} Non-overlapping frequency priors.}
\begin{ruledtabular}
\begin{tabular}{ccc}
Param. & Prior type &Prior range\\
 \hline
 \\
 $f_{*1}$&   LogUniform  & $(10^{-3}$ Hz$,10^{1.4}$ Hz$)$\\
 $f_{*2}$&   LogUniform  & $(10^{1.7}$ Hz$,10^{3}$ Hz$)$\\
\end{tabular}
\end{ruledtabular}
\end{table}

The results for this run are presented in Fig. \ref{fig:corner_non} in appendix A.

Next we turned to a more restrictive search, investigating LVK data's capacity of excluding certain zones of the parameter space. First, we fix the location of the second peak outside the sensitivity range $f_{*2} = 300$ Hz, and we select three values of the location of the first peak past the lower bound of the sensitivity range. Second, we perform the opposite search: we fix the location of the first peak, before the detector's sensitivity $f_{*1} = 10$ Hz, and we choose three values for the location of the second peak, past the peak sensitivity range. For consistency, the priors assigned to the remaining free parameters $\Omega_{ref}, \Omega_{*1}, \Omega_{*2}, n_2, n_3 $ were the same as those adopted in the earlier analysis. The results can be seen in Fig. \ref{fig:n2} and Fig. \ref{fig:n3}, respectively.

In all these situations, the highest value for the Bayes factor between a model containing only noise and the model composed of the \textit{CBC + Cosmological Double Peak} gravitational waves stochastic background has the value: 
\[logB^{CBC+Cosmo}_{Noise} = -0.97\]

No detection of such a background can be claimed. However, the posterior distributions show that specific $(n_2, \Omega_{*1})$ and $(n_3, \Omega_{*2})$ pairs are inconsistent with the observational results. A more detailed discussion of this point is presented in Section V. 

A complementary analysis was performed to illustrate the data’s sensitivity to double-peaked spectra. In this approach, the amplitudes of the two peaks were fixed to be equal, while the frequency of the second peak was expressed as a function of the first peak via the relation:
\[f_{*2} = R \times f_{*1}\]
The priors for this analysis are given in Table \ref{table:tab3}.
\begin{table}[h!]
\caption{\label{table:tab3}Priors selected for the equal peak amplitude search.}
\begin{ruledtabular}
\begin{tabular}{ccc}
Param. & Prior type &Prior range\\
 \hline
 \\
 $\Omega_{ref}$   & LogUniform    &$(10^{-11},10^{-6})$\\
$\Omega_{*1}$&   LogUniform  & $(10^{-11},10^{-6})$ \\
$f_{*1}$&   LogUniform  & $(10^{-3}$ Hz$,10^{3}$ Hz$)$\\
 $n_2$& Uniform  &$(-6, 0)$ \\
 $n_3$& Uniform  &$(0, 6)$ \\
\end{tabular}
\end{ruledtabular}
\end{table}

The analysis was run for 4 different numerical values of $R$: $10$, $50$, $100$, and $500$, and the results are presented in Fig. \ref{fig:R_varies} in appendix A.

The data sensitivity manifests as a distinct double-dip structure in the $f_{*1}$ vs $\Omega_{*1}$ parameter space, becoming more pronounced as the value of $R$ is increasing. However, for $R=10$ we obtain only a single-dip structure. This serves as a validation of the analysis, as it reproduces to a good approximation the same posterior plots for the single-peak case. We recover the $(n_2, \Omega_{*1})$ and $(n_3, \Omega_{*1})$ exclusion zones, when the spectral peaks are outside the detector's sensitivity.\footnote{Physically the condition would involve $\Omega_{*2}$, but here the two peak amplitudes are equal $\Omega_{*1}=\Omega_{*2}$.}

\section{Results}

The wide priors search (Fig. \ref{fig:corner_wide}) returns flat posteriors for most slope parameters. However, $n_2$ (the fall after the first peak) and $n_3$ (the rise into the second peak) show amplitude–slope correlations in the following parameter spaces: $n_2$ vs. $\Omega_{*1}$ and $n_3$ vs. $\Omega_{*2}$. The key qualitative trend is:
\begin{itemize}
    \item The data disfavor gently varying inter-peak regions at large peak amplitudes. Conversely, for spectra characterized by larger values of $n_3$ or smaller values of $n_2$—corresponding to sharper transitions between the peaks—the valley lies below the detector’s sensitivity band, and thus no meaningful constraints can be placed, even at higher peak amplitudes. In this figure, for example, we notice that for $\Omega_{*1} = 10^{-7}$, only values of $n_2 \le -1$ are acceptable, and for $\Omega_{*2} = 10^{-7}$, only values of $n_3\ge2$ are allowed. 
    \item Enforcing non-overlapping peak frequencies priors sharpens the same effect (Fig. \ref{fig:corner_non}). 
    \item The search with fixed peak frequency ratios $R$ (Fig. \ref{fig:R_varies}) also exhibits sensitivity to  $n_2$ and $n_3$.
\end{itemize}

This underlines that the valley morphology between the peaks—set mainly by $n_2$ and $n_3$—is what the LVK band constrains best for these frequency priors.

Figs. \ref{fig:n2} and \ref{fig:n3} make the trend clearer by pinning one peak outside the band and scanning the other. As the movable peak approaches the sensitive band, the excluded region in the corresponding $(n, \Omega_{*})$ corner plot grows out of the mild slopes range and expands toward sharper slopes at high $\Omega_*$. 

When the first peak ($f_{*1}$) is shifted from sub-hertz values toward the observable range while the second peak remains fixed at $300\,\mathrm{Hz}$ (Fig. ~\ref{fig:n2}), the posterior in the $(n_2,\Omega_{*1})$ plane progressively narrows. At low $f_{*1}$, the distribution is broad and prior-dominated, indicating little sensitivity to either parameter. As $f_{*1}$ approaches the detector’s most responsive region (tens of hertz), large amplitudes $\Omega_{*1}$ become increasingly incompatible with moderate $n_2$ values, and the exclusion contours expand toward smaller $n_2$, revealing the data’s preference against slowly decaying post-peak slopes at high amplitude.

A complementary behaviour appears in Fig. ~\ref{fig:n3}, where the first peak is fixed below the band ($f_{*1} = 10\,\mathrm{Hz}$) and the second peak is moved from $200$ to $400\,\mathrm{Hz}$. As $f_{*2}$ is pushed beyond the sensitive band, the posterior in the $(n_3,\Omega_{*2})$ plane broadens and eventually flattens, demonstrating loss of constraining power. When $f_{*2}$ lies near the band center, however, the data exclude combinations of large $\Omega_{*2}$ with small or moderate $n_3$, corresponding to gently rising transitions into the second peak. 

In both cases, the trend indicates that broad inter-peak slopes are constrained when they
fall within the detector’s sensitivity, while sharper transitions (large $n_3$, small $n_2$) shift the valley below the sensitivity curve, leaving those regions of parameter space effectively unconstrained.

\section{Conclusions}

We performed a dedicated search for stochastic gravitational-wave backgrounds with a double-peaked spectral morphology using data from the LIGO--Virgo--KAGRA observing runs O1 through O4a. No statistically significant evidence for such a signal was found, and all results remain consistent with Gaussian noise. Nevertheless, the analysis provides direct constraints on the morphology of potential multi-peak spectra within the ground-based detection band. In particular, correlations observed in the posterior distributions between the peak amplitudes and their spectral slopes demonstrate that the current data can exclude broad inter-peak regions at large amplitudes, while steep, sharply separated peaks remain unconstrained when their in-between valley lies below the detector sensitivity. These results confirm that the LVK network already reaches the sensitivity necessary to probe complex, non-power-law stochastic backgrounds and can meaningfully constrain their shape even in the absence of a clear detection.

Looking ahead, the methodology developed here establishes a framework for future targeted searches for structured stochastic backgrounds. The continuation of the O4 run and the upcoming O5 run will significantly enhance sensitivity in the most constraining frequency band identified in this study. Next-generation observatories such as the Einstein Telescope \cite{Punturo_2010, Branchesi_2023, ET:2025xjr} and Cosmic Explorer \cite{Reitze:2019iox, galaxies10040090}, with broader bandwidths and improved low-frequency reach, will be capable of resolving multi-peak features that are inaccessible to current detectors. In combination with space-based missions like LISA \cite{Bartolo_2022, Bartolo_2016, causality, Caprini2016, auclair2020probing}, these experiments will enable a fully multi-band exploration of cosmological sources such as multi-step phase transitions, scalar-induced gravitational waves, and topological defect networks.

The present work thus provides an essential step toward detecting or constraining non-trivial spectral structures in the stochastic gravitational-wave background and toward uncovering the physical mechanisms that shaped the early Universe.

\section*{Acknowledgments}
The authors would like to express their gratitude to Fabrizio Rompineve for fruitful discussions during the development of this project and to Alba Romero for her valuable comments and suggestions on this manuscript. The authors acknowledge financial support from the Spanish Ministry of Science and Innovation (MICINN) through the Spanish State Research Agency, under Severo Ochoa Centres of Excellence Programme 2025-2029 (CEX2024001442-S). This work is partially supported by the Spanish MCIN/AEI/10.13039/501100011033 under the Grants  No. PGC2018-101858-B-I00, No. PID2020-113701GB-I00 and No. PID2023-146686NB-C31, some of which include ERDF funds from the European Union, and by the MICINN with funding from the European Union NextGenerationEU (PRTR-C17.I1) and by the Generalitat de Catalunya. C.-A. Miritescu received support from H2020-MSCA-RISE-2020 PROBES action (GA - 101003460). IFAE is partially funded by the CERCA program of the Generalitat de Catalunya. This material is based upon work supported by NSF's LIGO Laboratory which is a major facility fully funded by the National Science Foundation.
\clearpage

\twocolumngrid
\cleardoublepage
\newpage

\bibliography{bibliography}
\newpage
\onecolumngrid
\appendix
\renewcommand{\thefigure}{A\arabic{figure}}
\setcounter{figure}{0}
\section{Additional Figures}
\begin{figure*}[!b]
\centering
\includegraphics[width =1\linewidth]{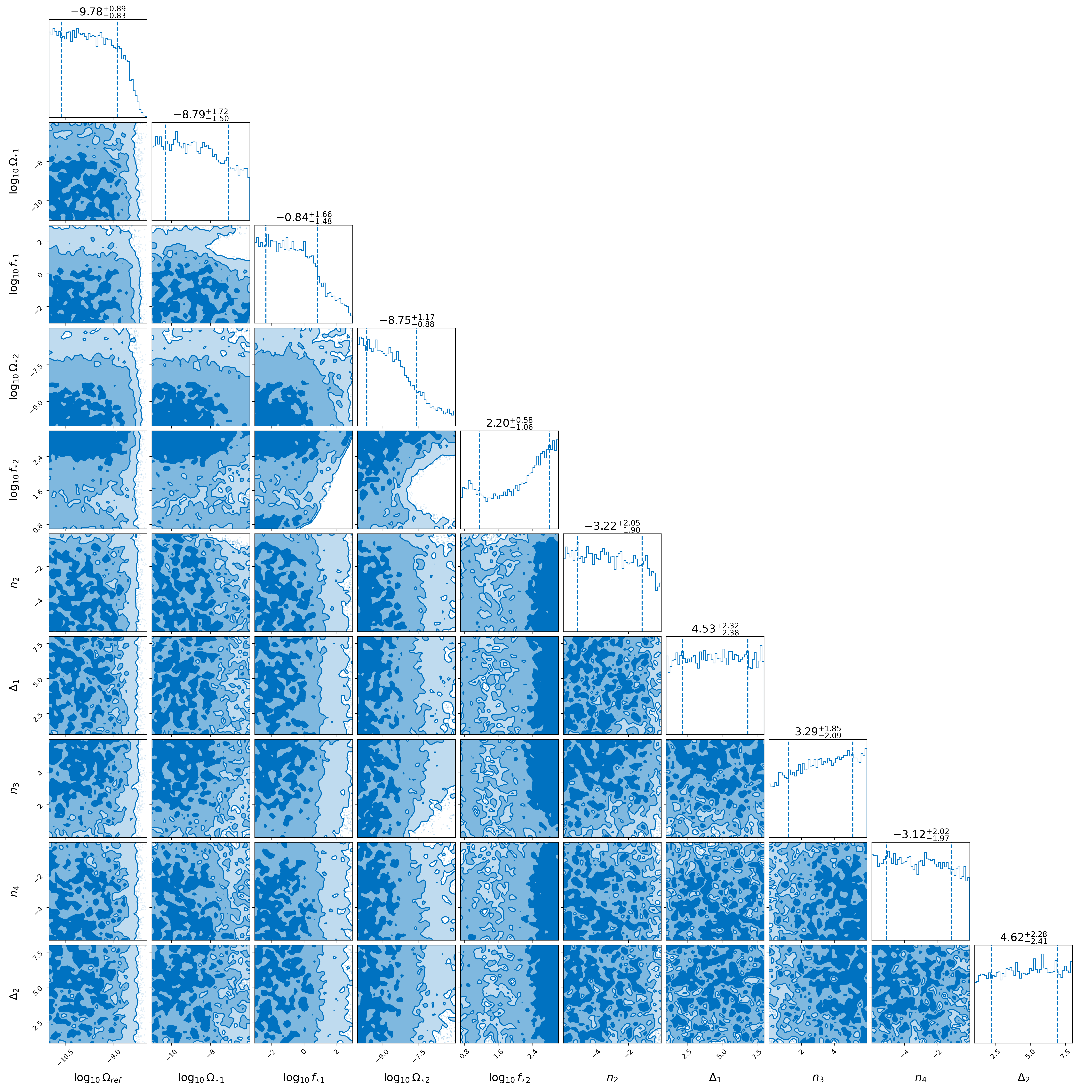}
\caption{Corner plot results for the Bayesian search using the parameters described above.}
\label{fig:corner_wide}
\end{figure*}

\begin{figure*}
\centering
\includegraphics[width =1\linewidth]{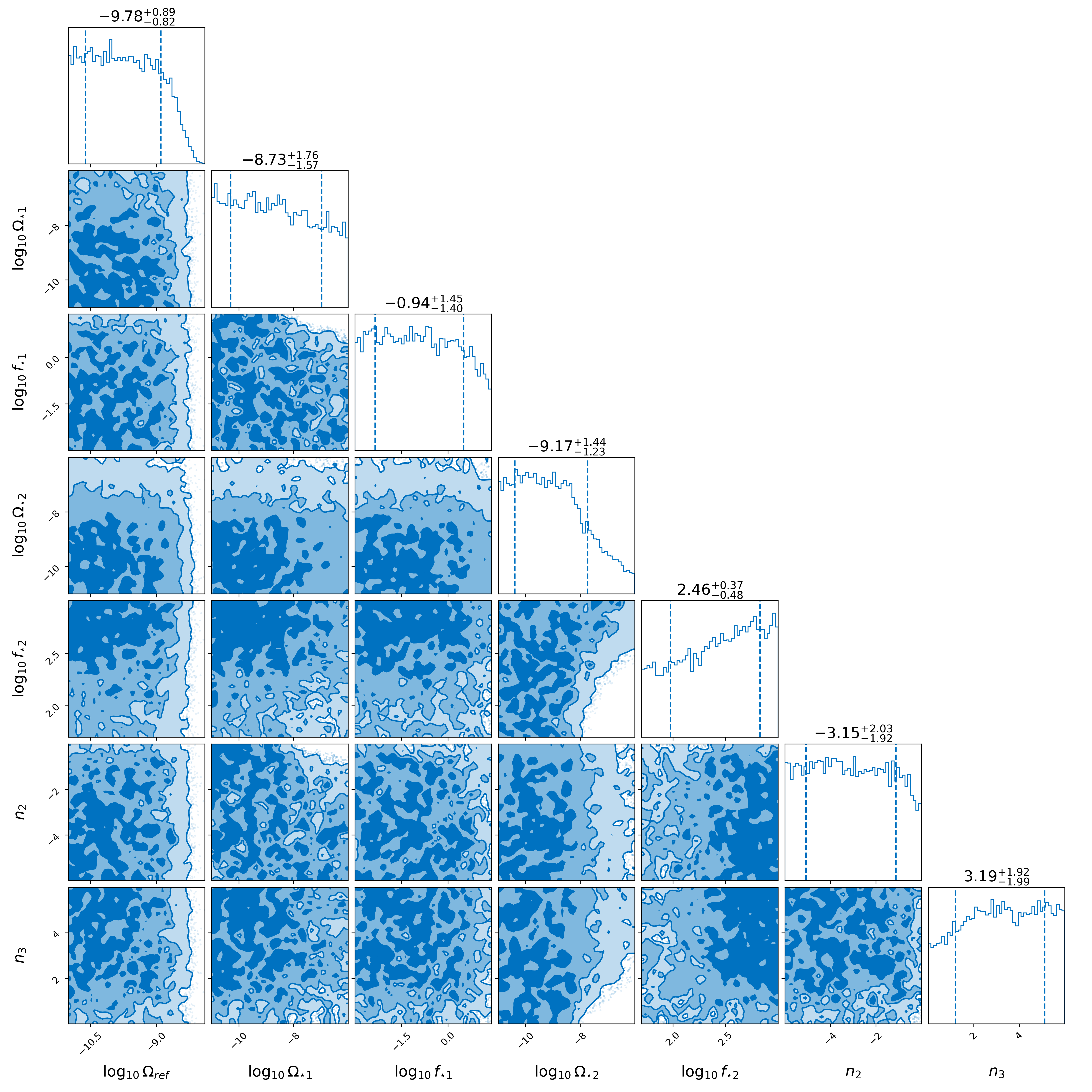}
\caption{Corner plot results for the Bayesian search using the non-overlapping frequency priors.}
\label{fig:corner_non}
\end{figure*}

\begin{figure*}
    \centering
    \begin{subfigure}[t]{0.48\textwidth}
        \centering
        \includegraphics[width=\linewidth]{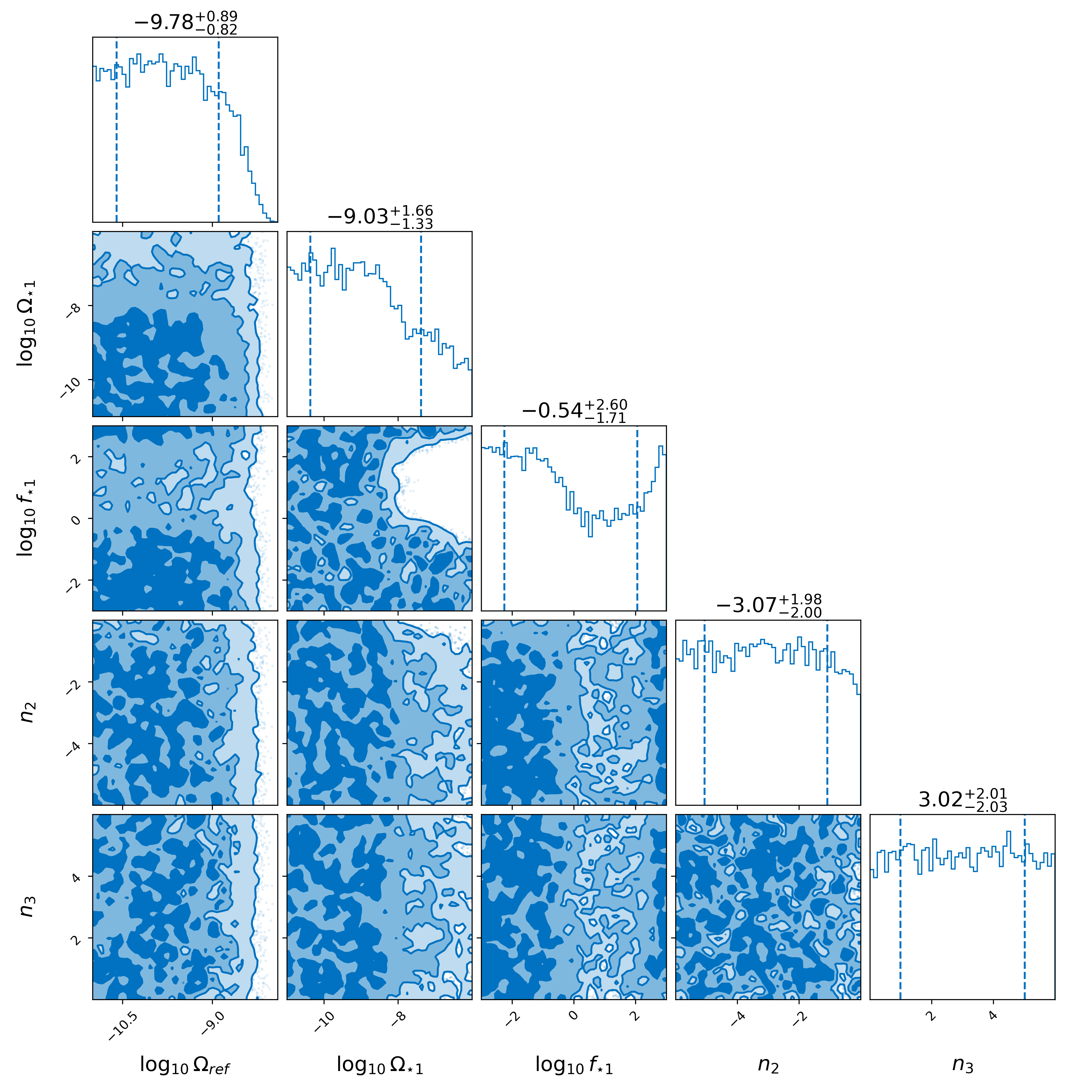} 
        \caption{$R=10$} \label{fig:R10}
    \end{subfigure}
    \hfill
    \begin{subfigure}[t]{0.48\textwidth}
        \centering
        \includegraphics[width=\linewidth]{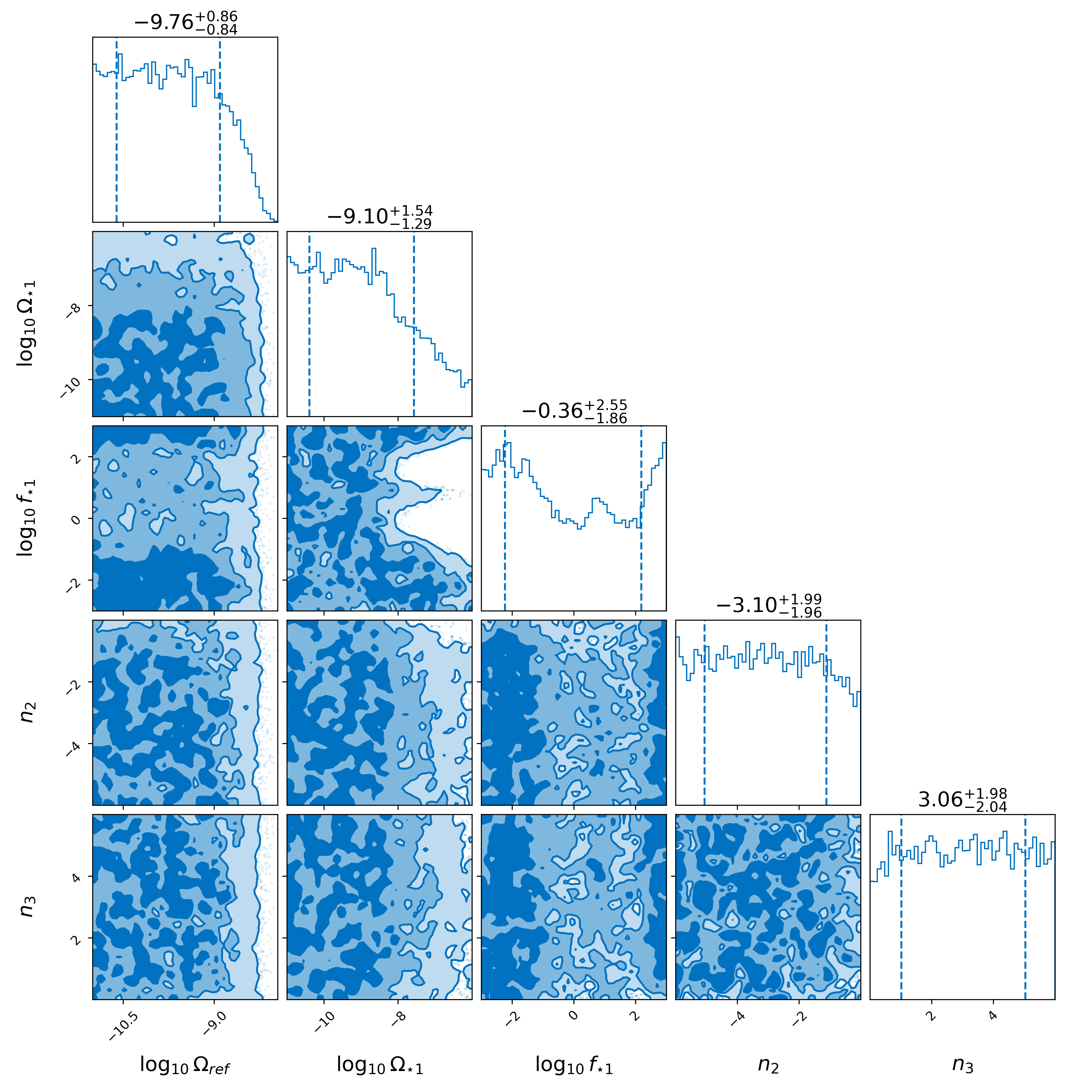} 
        \caption{$R=50$} \label{fig:R50}
    \end{subfigure}
    \begin{subfigure}[t]{0.48\textwidth}
    \centering
        \includegraphics[width=\linewidth]{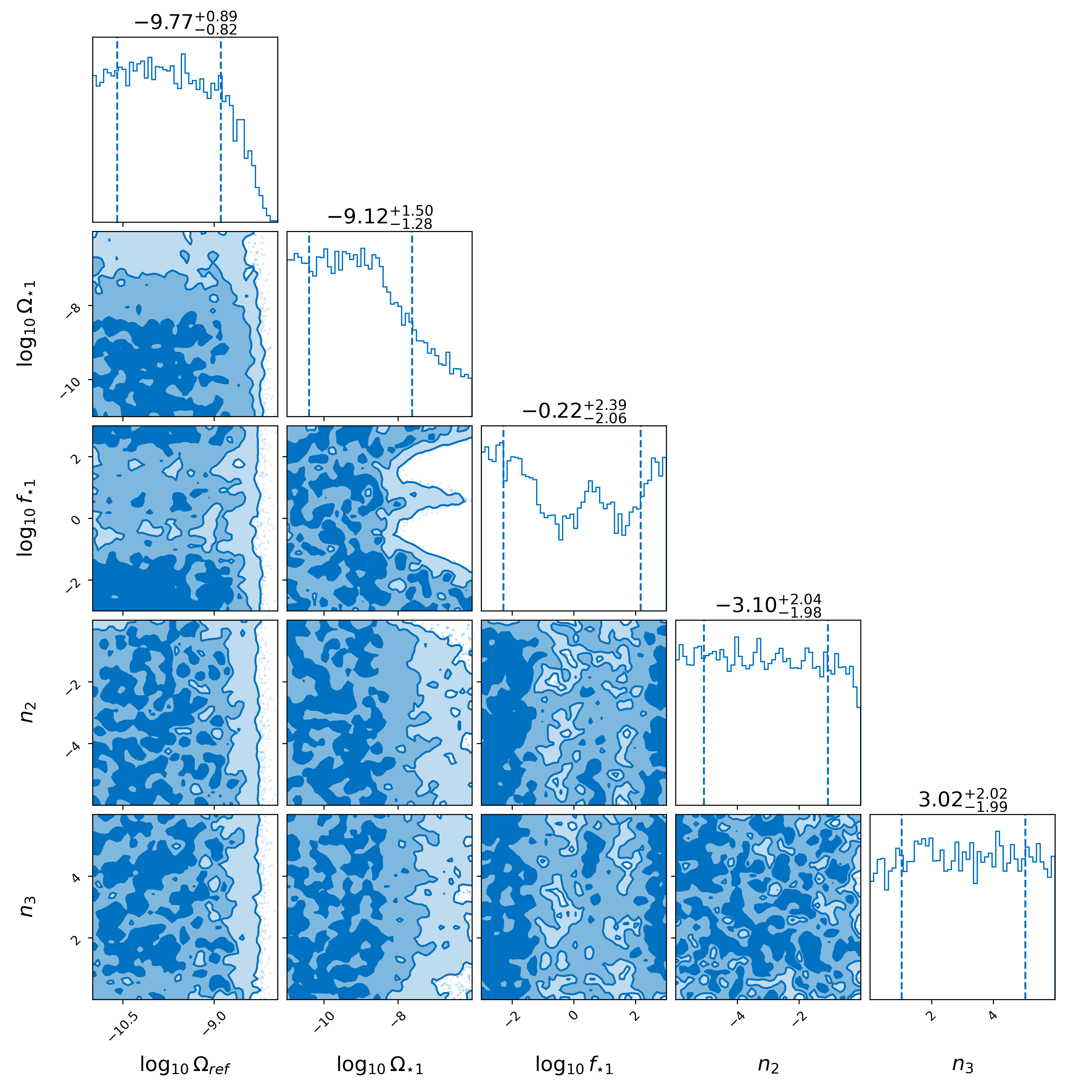} 
        \caption{$R=100$} \label{fig:R100}
    \end{subfigure}
    \hfill
    \begin{subfigure}[t]{0.48\textwidth}
    \centering
        \includegraphics[width=\linewidth]{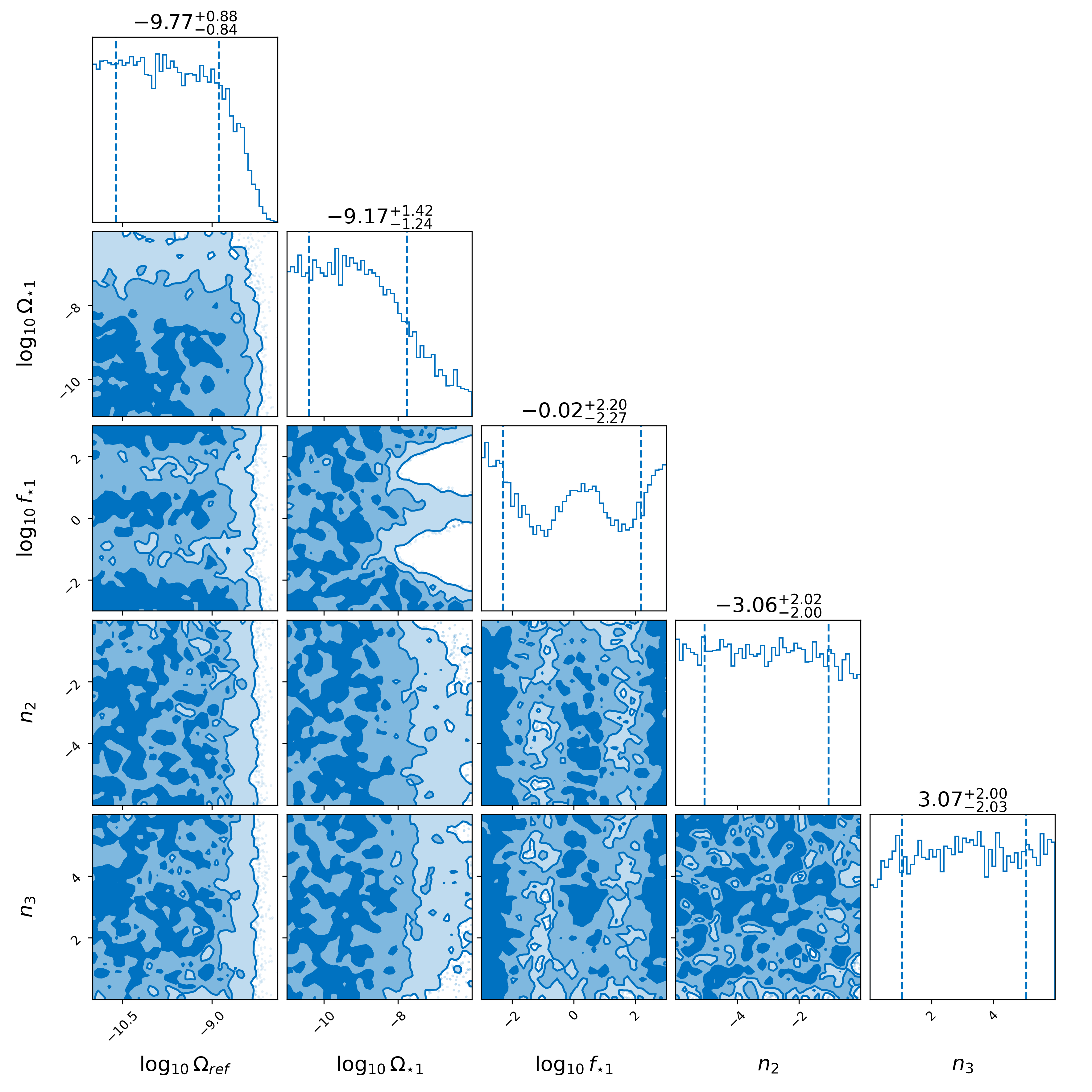} 
        \caption{$R=500$} \label{fig:R500}
    \end{subfigure}
    \caption{Equal amplitude ($\Omega_{*1}=\Omega_{*2}$) searches, with  different peak-frequencies ratios $R =f_{*2}/f_{*1}$. } \label{fig:R_varies}
\end{figure*}

In Appendix A we present additional figures supporting the main analysis. These include full corner plots of all sampled parameters for the wide-prior search (Fig. \ref{fig:corner_wide}), the scan enforcing non-overlapping peak-frequency priors (Fig. \ref{fig:corner_non}), and the equal-amplitude searches across several peak-frequency ratios (Fig. \ref{fig:R_varies}), providing a detailed view of parameter correlations and exclusion regions.

\end{document}